\documentclass[conference,a4paper]{APSIPA2025}
\usepackage{amsmath}
\usepackage{graphicx}
\usepackage{multirow}
\usepackage{threeparttable}
\usepackage[backend=biber,style=ieee,]{biblatex}
\usepackage{geometry}
\usepackage{multirow}
\usepackage{xurl}
\usepackage{url}
\usepackage{graphicx}
\usepackage{subfigure}
\usepackage{breqn}
\usepackage{amsmath}
\usepackage{mathtools}

\usepackage{caption}
\usepackage{fancyhdr}

\fancypagestyle{firststyle}{
  \fancyhf{}
  \fancyhead[C]{2025 Asia Pacific Signal and Information Processing Association Annual Summit and Conference (APSIPA ASC)}
}

\begin{document}

\title{Investigation of perception inconsistency in speaker embedding for asynchronous voice anonymization}

\author{
\authorblockN{
Rui Wang\authorrefmark{1},
Liping Chen\authorrefmark{1}, Kong Aik Lee\authorrefmark{2}, 
Zhengpeng Zha\authorrefmark{1},
Zhenhua Ling\authorrefmark{1}}

\authorblockA{
\authorrefmark{1}
University of Science and Technology of China \\
E-mail: wangrui256@mail.ustc.edu.cn  \{lipchen, zhazp, zhling\}@ustc.edu.cn}

\authorblockA{
\authorrefmark{2}
The Hong Kong Polytechnic University \\
E-mail: kong-aik.lee@polyu.edu.hk}
}

\maketitle
\thispagestyle{firststyle}
\pagestyle{fancy}

\begin{abstract}
Given the speech generation framework that represents the speaker attribute with an embedding vector, asynchronous voice anonymization can be achieved by modifying the speaker embedding derived from the original speech. However, the inconsistency between machine and human perceptions of the speaker attribute within the speaker embedding remains unexplored, limiting its performance in asynchronous voice anonymization. To this end, this study investigates this inconsistency via modifications to speaker embedding in the speech generation process. Experiments conducted on the FACodec and Diff-HierVC speech generation models discover a subspace whose removal alters machine perception while preserving its human perception of the speaker attribute in the generated speech. With these findings, an asynchronous voice anonymization is developed, achieving 100\% human perception preservation rate while obscuring the machine perception. Audio samples can be found in \url{https://voiceprivacy.github.io/speaker-embedding-eigen-decomposition/}.
\end{abstract}
\let\thefootnote\relax\footnotetext{\emph{Corresponding author: Liping Chen}.

This work was supported in part by the National Key Research and Development Program Project 2024YFE0217200, the Innovation and Technology Fund of the Hong Kong SAR MHP/048/24, the National Natural Science Foundation of China under Grant U23B2053, and the Fundamental Research Funds for the Central Universities WK2100000043.}
\section{Introduction}
Advancements in speech technologies have intensified security risks related to the misuse of speaker attributes, necessitating the development of voice privacy protection techniques. In this context, voice anonymization, originating from the 1980s \cite{ref121}, has regained the interest of the community as it provides a viable solution to protecting speaker attributes from being extracted by speaker models. To date, voice anonymization can be realized in both synchronous\cite{fang2019speaker,2020Introducing,tomashenko2022voiceprivacy} and asynchronous\cite{chen2023voicecloak,deng2023vCloak,wang2024async} manners. Synchronous voice anonymization alters both machine-discernible and human-perceivable speaker attributes. Asynchronous voice anonymization modifies the machine-discernible attribute while preserving the human-perceivable attribute.

Facilitated by the speech generation framework wherein the speaker attribute is disentangled and represented with an embedding, voice anonymization can be realized by replacing the original speaker with a pseudo-speaker \cite{fang2019speaker,wang2024async}. In asynchronous voice anonymization, the construction of the pseudo-speaker forms a critical challenge. An asynchronous voice anonymization method was proposed in \cite{wang2024async}, where the pseudo-speaker embedding was generated by introducing adversarial perturbation to the speaker embedding. In this approach, stronger perturbations provided better protection against machine perception with larger alteration of the machine-discernible speaker attribute,  whereas weaker perturbation better preserves the human-perceivable attribute. Due to the lack of differentiating between machine and human perceptions in the speaker embedding, the adversarial method exhibits limitations in protecting machine perception of the speaker attribute while preserving its human perception.

This paper investigates the differences between machine and human perceptions of speaker attributes. Given a speech generation model that incorporates speaker attribute disentanglement and its representation via an embedding vector, speaker-modified speech utterances are generated through modifications to the speaker embedding. Specifically, the modification is performed in a subspace of the speaker embedding by eliminating its contribution from the embedding. Machine and human perceptions of the speaker attribute within the speaker-modified utterances were examined individually in the experiments. To our knowledge, this is the first investigation of the differences between the machine and human perceptions within the speaker embedding in the context of speech generation. Our contributions are as follows:

\begin{enumerate}
\item Experimental findings in two speech generation models, FACodec\cite{FAcodec} and Diff-HierVC\cite{choi2023diff}, demonstrate that a speaker variability subspace exists, whose removal exclusively influences machine perception of the speaker attribute without affecting its human perception. This reveals the inconsistency between machine and human perceptions of speaker attributes within the speaker embedding.

\item An asynchronous voice anonymization method is developed, in which the pseudo-speaker vector is obtained by removing the subspace contribution from the original speaker vector. It achieves a 100\% human perception preservation rate while obscuring the machine-discernible speaker attributes in the anonymized speech.
\end{enumerate}

 \begin{figure}[!t]
        \centering
        \includegraphics[scale=1.00]{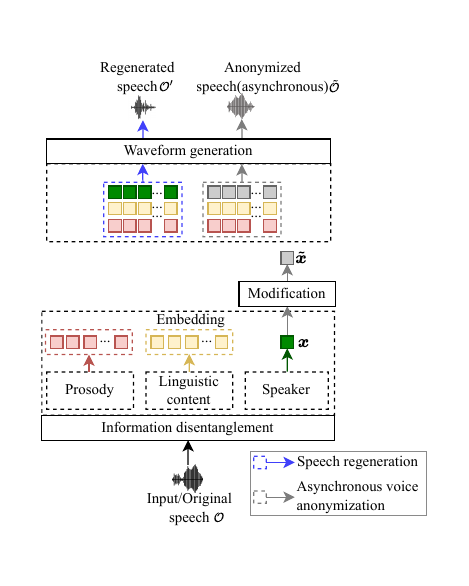}
        \caption{Speech generation framework based on information disentanglement. Colored boxes represent embeddings of disentangled attributes: red (prosody), yellow (linguistic content), green (original speaker $\boldsymbol{x}$). The gray box represents the modified speaker embedding $\tilde{\boldsymbol{x}}$. The blue and gray rectangular boxes with dotted lines, along with the corresponding arrowed lines, indicate speech regeneration and speaker-modified speech generation processes, respectively.}
        \label{fig: architecture}
\end{figure}

\section{Background}
\label{sec: background}
Fig. \ref{fig: architecture} illustrates the framework for information disentanglement, facilitating the generation of speaker-modified speech. The disentanglement-based speech generation framework and its application in generating speaker-modified speech are detailed in the following.
\subsection{Disentanglement-based speech generation}
Fig. \ref{fig: architecture} presents the speech generation framework based on information disentanglement and waveform generation. In the information disentanglement phase, given an input speech utterance ${\mathcal{O}}$, three distinct attributes, including prosody, linguistic content, and speaker characteristics, are disentangled and represented by separate embedding vectors. A specific configuration is illustrated where prosody and linguistic content attributes are extracted from speech frames and represented with sequences of embedding vectors. Besides, the speaker attribute is encoded for the entire utterance using a single vector ${\boldsymbol{x}}$. In the waveform generation phase, ${\boldsymbol{x}}$ is replicated to match the length of the prosody and content embedding sequences. The embedding vectors of the three attributes are input into the waveform generation module, producing a speech waveform $\mathcal{O}^{\prime}$, which is a regenerated version of ${\mathcal{O}}$. The disentanglement-based speech generation mechanism enables control over generated speech by facilitating the manipulation of speech attributes, especially prosody \cite{GST,zhao2022disentangling} and speaker characteristics\cite{casanova2022yourtts,choi2023diff,li2023freevc,FAcodec}.

\subsection{Anonymized speech (asynchronous) generation}
In the asynchronous voice anonymization method \cite{wang2024async} built upon the speech generation framework depicted in Fig. \ref{fig: architecture}, given an original utterance, the prosody, linguistic content, and speaker attributes were disentangled and represented with separate embedding vectors. The speaker embedding vector $\boldsymbol{x}$, denoted with the green square box, was subsequently modified to $\tilde{\boldsymbol{x}}$ as the pseudo-speaker vector (the gray square box in Fig. \ref{fig: architecture}). Thereby, the anonymized speech $\tilde{\mathcal{O}}$ was generated by the waveform generation module and used as the anonymized speech, utilizing the original prosody and linguistic content embedding vectors and the modified speaker vector ${\tilde{\boldsymbol{x}}}$.

\section{Speaker modification}

Given the speech generation framework depicted in Fig. \ref{fig: architecture}, our investigation into the differences between machine and human perceptions of speaker attributes in speaker embedding is conducted through modifications to the original speaker embedding ${\boldsymbol{x}}$, followed by experimental evaluations of both perceptions within the speaker-modified utterances \({\tilde{\mathcal{O}}}\). The modification is performed within the variability subspaces of speaker embedding as detailed in the following.



\subsection{Variability space}
A speaker embedding vector ${\boldsymbol{x}}$ is hypothesized to be decomposible with a basis matrix ${\boldsymbol{V}}$ as follows:

\begin{equation}
\label{eq: orthogonal decomposition}
    {\boldsymbol{x}}={\boldsymbol{V}}{\boldsymbol{c}}.
\end{equation}
Assume ${\boldsymbol{x}}$ to be a $D$-dimensional vector, ${\boldsymbol{V}}$ is composed of $D$ orthogonal unit vectors as ${\boldsymbol{V}}=\left[{\boldsymbol{v}}_1,...,{\boldsymbol{v}}_D\right]$, with the vectors $\left\{{\boldsymbol{v}}_1,...,{\boldsymbol{v}}_D\right\}$ serving as the \emph{basis vectors}. The vector $\boldsymbol{c} = \left[c_{1}, ..., c_{D}\right]^{\mathsf{T}}$ is the coefficient vector, with each element quantifying the contribution of the corresponding basis vector in constructing ${\boldsymbol{x}}$. The space spon by $\boldsymbol{V}$ captures the variability of $\boldsymbol{x}$, and is referred to as \emph{variability space}.

Given $N$ speech utterances, the speaker embedding vectors are extracted and denoted as $\mathcal{X}=\cup_{n=1}^N{\boldsymbol{x}}_n$. Firstly, the covariance matrix ${\boldsymbol{\Sigma}}$ is calculated from ${\mathcal{X}}$.

Thereafter, ${\boldsymbol{V}}$ is derived via eigen-decomposition of ${\boldsymbol{\Sigma}}$ as follows:
\begin{equation}
    \boldsymbol{\Sigma} = \boldsymbol{V} \boldsymbol{\Lambda} \boldsymbol{V}^{\mathsf T},
\end{equation}
where $\boldsymbol{V} = \left[ \boldsymbol{v}_1,\boldsymbol{v}_2,\dots,\boldsymbol{v}_D \right]$ is obtained as the matrix of eigenvectors, and $\boldsymbol{\Lambda} = \text{diag}(\lambda_1, \lambda_2, \dots, \lambda_d)$ is a diagonal matrix with eigenvalues on the diagonal. Particularly, the eigenvalues are sorted in descending order as $\lambda_1 \geq \lambda_2 \geq \dots\geq\lambda_D$. An identical mechanism for variability space derivation is utilized in principal component analysis (PCA) \cite{PRML}. Readers are referred to it for mathematical details.

\subsection{Speaker embedding modification}

Given the speaker vector ${\boldsymbol{x}}_n$ extracted from the $n$-th utterance, the coefficient vector in space ${\boldsymbol{V}}$ is obtained as follows:

\begin{equation}
    {\boldsymbol{c}}_n={\boldsymbol{V}}^{\mathsf T}{\boldsymbol{x}}_n,
\end{equation}
where ${\boldsymbol{c}}_n$ consists of $D$ elements as $\left\{c_{n,1},...,c_{n,D}\right\}$. The modification of ${\boldsymbol{x}}_n$ is achieved by altering ${\boldsymbol{c}}$ to $\tilde{\boldsymbol{c}}_{n}$ which is conducted on the subspaces of ${\boldsymbol{V}}$. Specifically, a subspace is characterized by three parameters: initial dimension $i$, subspace size $ K $, and the span direction (forward or backward). It is represented as ${\boldsymbol{S}}=\left\{i,K, \rm {direction}\right\}$, indicating that the subspace spans $K$ dimensions from the $i$-th dimension in the designated direction. The contribution of ${\boldsymbol{S}}$ is eliminated from \( {\boldsymbol{x}} \) by setting the coefficients associated with its basis vectors to 0 as follows:
\begin{equation}
\begin{aligned}
&c_{n,i},...,c_{n,i+K-1}=0, & \text{if direction = +} \\
&c_{n,i-K+1},...,c_{n,i}=0, & \text{if direction = -}
\end{aligned},
\end{equation}
where ``+'' (forward) and ``-'' (backward) are the span direction of the subspace. Combined with the remaining coefficients, the modified coefficient vector ${\tilde{\boldsymbol{c}}}_n$ is obtained. Finally, the modified speaker embedding vector $\tilde{\boldsymbol{x}}_n$ is obtained as follows:
\begin{equation}
\label{eq: orthogonal decomposition}
    {\tilde{\boldsymbol{x}}}_n={\boldsymbol{V}}\tilde{{\boldsymbol{c}}}_n.
\end{equation}

\begin{figure}[t]
        \centering
        \begin{tabular}{cccc}
            \subfigure[Sorted logarithmic eigenvalues]
            {\includegraphics[width=0.21\textwidth]{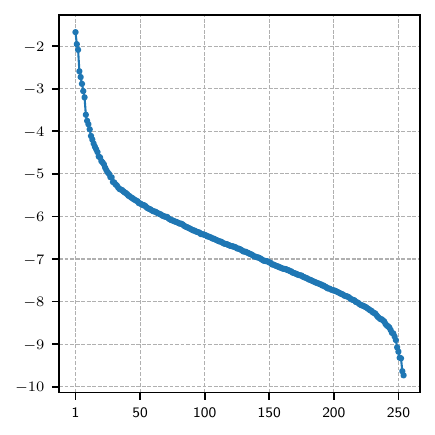}
            \label{fig: eigenvalue_a}
            } 
            
            & 
            \subfigure[Differentials (i.e., delta) of the sorted logarithmic eigenvalues]        
            {\includegraphics[width=0.21\textwidth]{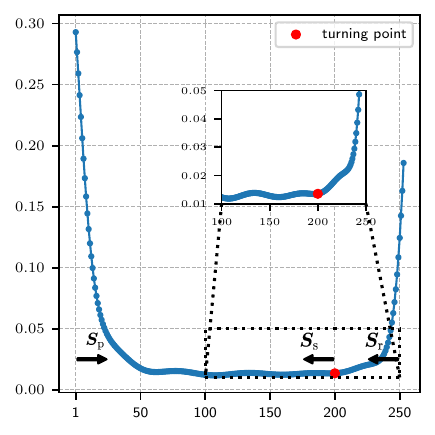}
            \label{fig: eigenvalue_b}
            } & 
        \end{tabular}
        \caption{Sorted logarithmic eigenvalues and their differentials of the speaker variability space within the open-source FACodec model. The speaker embedding is of dimension $D$=256. The arrows in Fig. \ref{fig: eigenvalue_b} indicate the three modification subspaces ${\boldsymbol{S}}_{\rm p}$, ${\boldsymbol{S}}_{\rm s}$ and ${\boldsymbol{S}}_{\rm r}$, starting from their initial dimensions and directing along their spans, respectively.}
        \label{fig: eigenvalue}
\end{figure}

\subsection{Modification subspaces}
For presentational clarity, the variability space of the speaker embedding in the open-source FACodec model\footnote{\label{facodecurl}\url{https://huggingface.co/spaces/amphion/naturalspeech3_facodec}} is adopted for description. Derived from the speaker embedding vector set extracted from the speech utterances in the LibriSpeech train-clean-360\cite{librispeech} dataset, the logarithmic eigenvalues, i.e., $\left\{log{\lambda}_1,...,log{\lambda}_D\right\}$, are plotted in Fig. \ref{fig: eigenvalue_a}. 
Fig. \ref{fig: eigenvalue_b} illustrates the delta of the logarithmic eigenvalues, computed as follows:
\begin{equation}
    a_i={\rm log}\lambda_{i+1}-{\rm log}\lambda_{i},
\end{equation}
for $i=1,...,D-1$. Particularly, the FACodec model employs speaker vectors of dimension $D$=256. 

As shown by the arrows in Fig. \ref{fig: eigenvalue_b}, given the delta log-eigenvalue, three subspace region are investigated for the modification of the speaker embedding: primary, secondary, and residual subspaces. The \emph{primary subspace} starts from the 1st dimension and spans in the forward direction of size $K_{\rm p}$, denoted as $\boldsymbol{S}_{\rm p} = \left\{1, K_{\rm p}, +\right\}$. It is composed of the dominant basis vectors of the variability space. The \emph{secondary subspace} is defined at the turning point where the differential oscillates slightly before and increases monotonically after. The point is marked by the red dot in Fig. \ref{fig: eigenvalue_b} with the dimension $i_{\rm s}$. The secondary subspace starts from $i_{\rm s}$ and spans in the backward direction with size $K_{\rm s}$, denoted as ${\boldsymbol{S}}_{\rm s}=\left\{i_{\rm s},K_{\rm s},-\right\}$. The \emph{residual subspace} spans from the last dimension $D$ in the backward direction with size $ K_{\rm r} $, denoted as $ {\boldsymbol{S}}_{\rm r}=\{D, K_{\rm r}, -\} $. Between them, $ {\boldsymbol{S}}_{\rm r}$ represents the least important subspace in the variability space. The importance of $ {\boldsymbol{S}}_{\rm s} $ is higher than $ {\boldsymbol{S}}_{\rm r} $ while lower than $ {\boldsymbol{S}}_{\rm p} $. In these subspaces, the subscripts $_{\rm p}$, $_{\rm s}$, and $_{\rm r}$ are short for primary, secondary, and residual, respectively.

\begin{figure*}[!t]
    \centering
    \begin{tabular}{cccc}
        \subfigure[FACodec]{\label{subfig: FACodec}\includegraphics[width=0.85\textwidth]{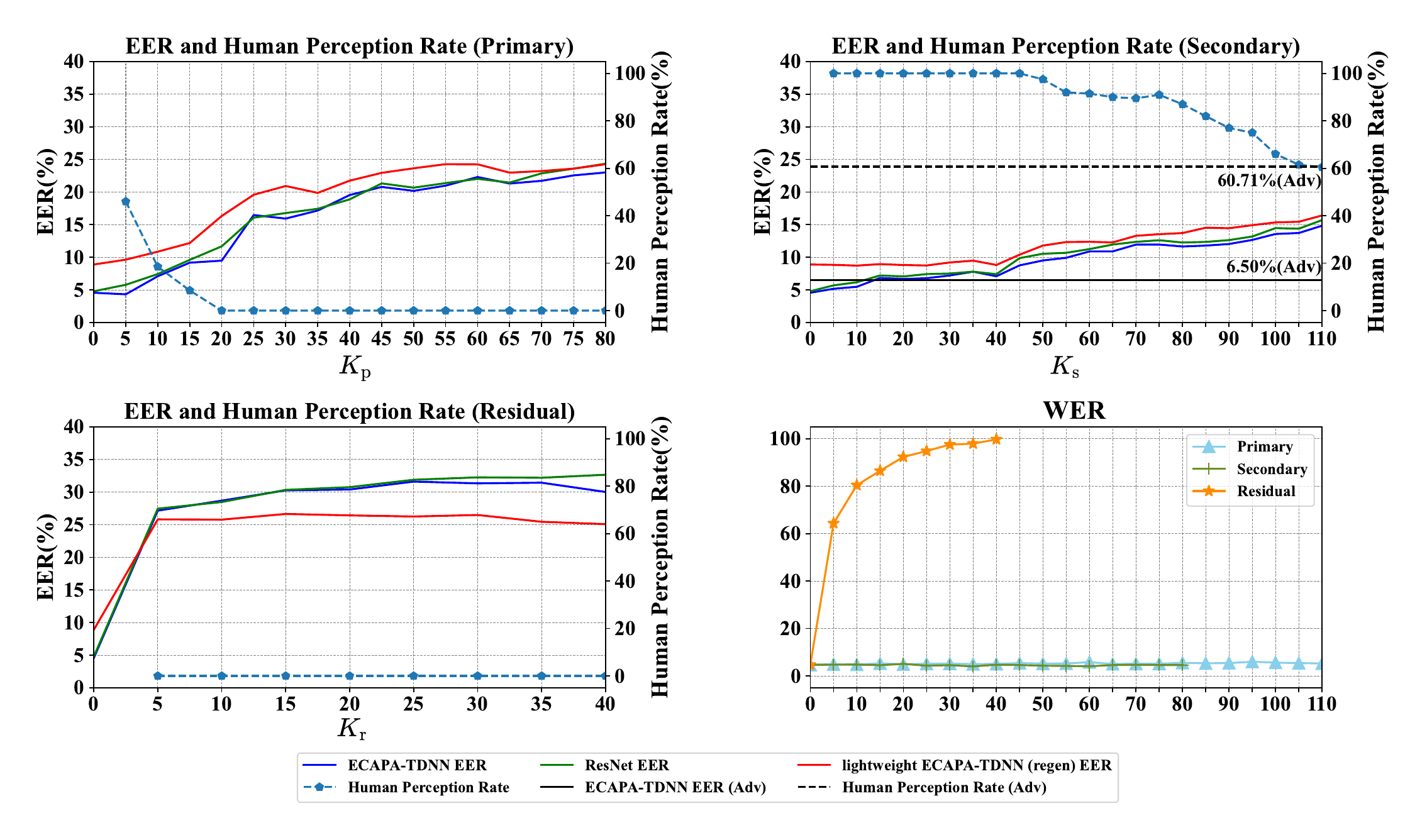}
        }
        \\
        
        \subfigure[Diff-HierVC]{\label{subfig: Diff-HierVC}\includegraphics[width=0.85\textwidth]{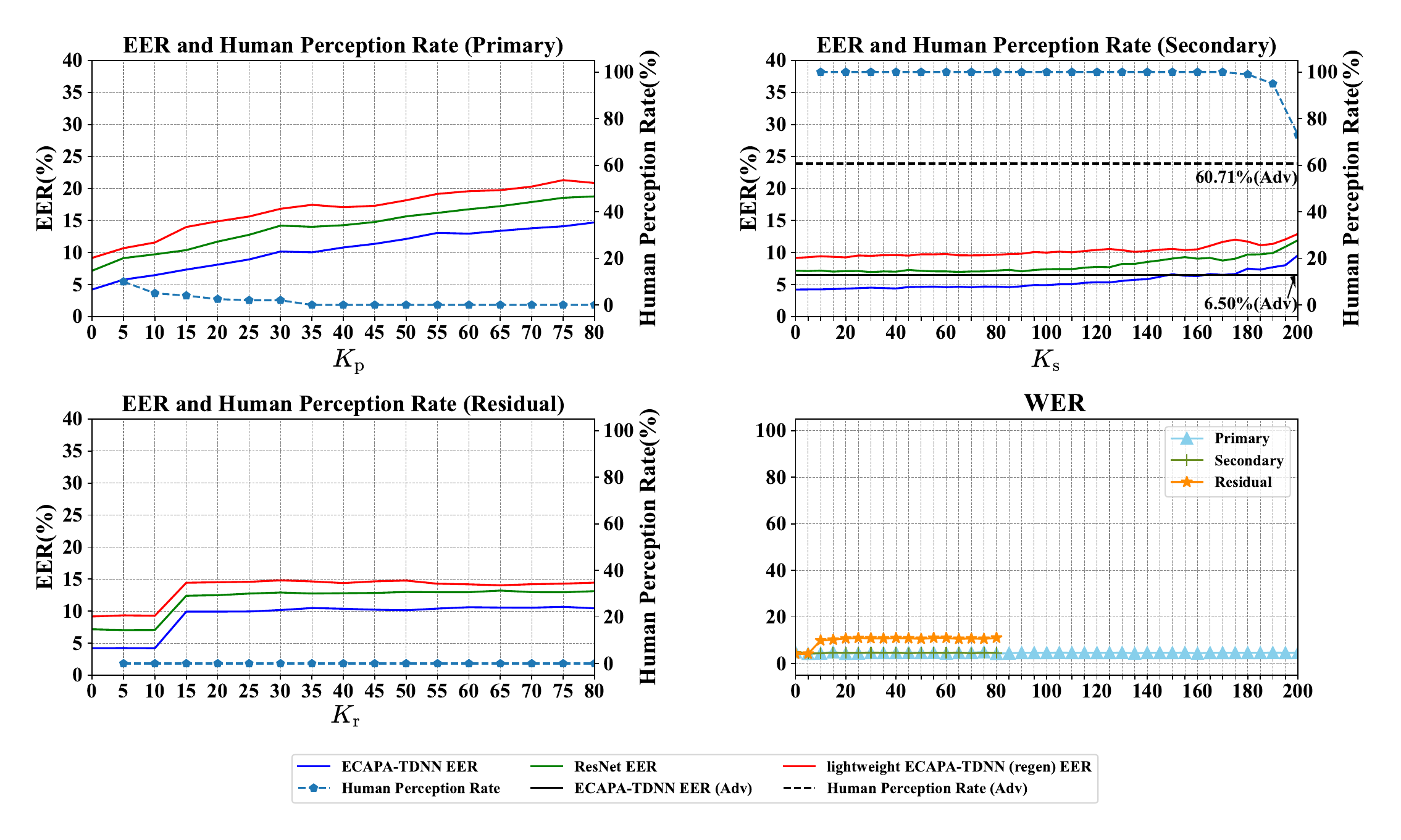}} & 
        \\
    \end{tabular}
    \caption{
       Machine perception, human perception, and linguistic content evaluation results across the primary, secondary, and residual subspaces under varying sizes. The results of FACodec and Diff-HierVC are shown in  Fig. \ref{subfig: FACodec} and Fig. \ref{subfig: Diff-HierVC}, respectively. 
       The machine perception (measured by  EER(\%)) and human perception (measured by human perception rate(\%)) of the primary and secondary subspace configurations are shown in the first row, in the first and second columns, respectively, while the results for the residual subspace are presented in the left column of the second row. The EERs obtained with the ECAPA-TDNN, ResNet, and lightweight ECAPA-TDNN trained with regenerated (regen) speech are included. The WERs(\%) obtained in the three configurations are presented in the right column of the second row. 
       The EER and human perception preservation rate obtained by the adversarial method (Adv) \cite{wang2024async} are included in the plots of secondary subspace with solid and dotted black lines, respectively.
    }
    \label{fig:example_grid}
    \vspace{-7pt}
\end{figure*}

\section{Experiments}
\label{sec: experiments}

\subsection{Dataset \& speech generation models}
Our evaluations were conducted on the dev-clean subset of the LibriSpeech \cite{librispeech} dataset, including 2,703 utterances from 20 female and 20 male speakers. All recordings were resampled to 16 kHz. The speaker embedding variability space was obtained from the LibriSpeech train-clean-360 dataset.
The open-source FACodec\footref{facodecurl} and Diff-HierVC models\footnote{\url{https://github.com/hayeong0/Diff-HierVC}} were examined as the speech generation model. The dimensions of the speaker vectors in both models are 256.
The turning dimensions $i_{\rm s}$ are 200 and 218 in FACodec and Diff-HierVC, respectively.


\subsection{Evaluation Metrics}
Evaluations were conducted to assess the machine and human perceptions of the speaker attributes in the utterances generated with the modified speaker embedding vectors. Given the degradation introduced by the speech generation model, the original utterance \( {\mathcal{O}} \) was regenerated using its extracted speaker vector $\boldsymbol{x}$, giving \( {\mathcal{O}^{\prime}} \). It served as the reference for a fair comparison with the speaker-modified speech \( {\tilde{\mathcal{O}}} \). Additionally, in line with the requirement of the voice anonymization task \cite{2020Introducing,tomashenko2022voiceprivacy,tomashenko2024voiceprivacy}, the linguistic content preservation capability was measured.

\begin{itemize}
    \item {\emph{Machine perception modification}:} Automatic speaker verification (ASV) evaluation was conducted to assess the modification of machine-perceivable speaker attributes. Three speaker embedding extractors were employed: ECAPA-TDNN\cite{desplanques2020ecapa}, ResNet\cite{resnet}, and a lightweight ECAPA-TDNN. Specifically, the ECAPA-TDNN and ResNet extractors were trained with the VoxCeleb1 \& 2 datasets\cite{VoxCeleb1,VoxCeleb2} using the ASVSubtools open-source toolkit\footnote{\label{subtools}\url{https://github.com/Snowdar/asv-subtools}}\cite{tong2021asv}. The lightweight ECAPA-TDNN extractor was trained on the regenerated speech of the LibriSpeech train-other-500 subset. The modified speech ${\tilde{\mathcal{O}}}$ was used for both enrollment and testing.
    Cosine similarities between speaker embeddings extracted by the three models were used for scoring. ASV performance was measured in terms of equal error rate (EER), where a higher EER indicates a stronger alteration of machine-discernible speaker characteristics. The ASV evaluations were conducted on the trials provided by VPC 2024\cite{tomashenko2024voiceprivacy}\footnote{\url{https://github.com/Voice-Privacy-Challenge/Voice-Privacy-Challenge-2024}}, with scores from male and female trials pooled together for EER calculation.
    \item {\emph{Human perception preservation}:} Subjective listening tests were conducted to assess the preservation of human perception. In each test, 200 utterances were randomly selected from the evaluation dataset. For each test utterance, given the pair of its regenerated version \( \mathcal{O}^{\prime} \) and the speaker-modified version \( \tilde{\mathcal{O}} \), five listeners were asked to decide whether the speakers were indistinguishable. Listeners gave a \emph{yes} (indistinguishable) or \emph{no} (distinguishable) for each utterance pair. A pair was decided to be perceived as the same speaker if it got a minimum of three yes.
    \item {\emph{Linguistic content preservation}:} The preservation of the linguistic content of the original speech was measured with automatic speech recognition (ASR) evaluations. The Whisper model\footnote{\url{https://platform.openai.com/docs/api-reference/introduction}}\cite{whisper} provided by OpenAI was called. The performances were measured with word error rates (WERs).
\end{itemize}




\subsection{Experimental configurations}
Our study investigated various subspaces by varying the sizes of the three subspace types. In the experiments on the primary subspace, the size $K_{\rm p}$ was examined from 0 to 80 with step 5 for both the FACodec and Diff-HierVC models.
In the secondary subspace experiments, the size $K_{\rm s}$ was examined from 0 to 80 with step 5 for the FACodec model. For the Diff-HierVC model, $K_{\rm s}$ was examined from 0 to 200 with step 5.
In the residual subspace experiments, size $K_{\rm r}$ was examined from 0 to 40 with step 5 for the FACodec model. For the Diff-HierVC model, $K_{\rm s}$ was examined from 0 to 80 with step 5.
Notably, in these experiments, the sizes of 0 indicate no modifications applied to the speaker embedding, resulting in the speech being the regenerated speech ${\mathcal{O}^{\prime}}$.
For comparison, the EER obtained in the ASV evaluation and the human perception preservation rate achieved by the adversarial method proposed in \cite{wang2024async} are presented together with the secondary subspaces, represented with \emph{Adv}. The ECAPA-TDNN speaker extractor was utilized in its ASV evaluation.

\subsection{Results}

In the primary subspace experiments, as ${K}_{\rm p}$ increased from 0, the ASV EERs rose rapidly to approximately 25\% on the three speaker extractors, i.e., ECAPA-TDNN, ResNet, lightweight ECAPA-TDNN. This indicates an obvious modification in the machine-discernible speaker attributes.
However, the human perception rates decreased sharply from 46.00\% to 0\% at $K_{\rm s}=20$, indicating its incapability of preserving the human perception of speaker attributes. These observations indicate that the removal of the contribution of the primary subspace alters both of speaker attributes in speaker-modified speech. This suggests that it is associated with both machine-discernible and human-perceptible speaker attributes.

The secondary subspace results show that the EER increased as the subspace size increased. Besides, in the FACodec model, the human perception rate remained at 100\% up to $K_{\rm s}=45$. Similar results are observed in the Diff-HierVC model with $K_{\rm s}$ ranging from 5 to 170. These observations demonstrate the removal of these secondary subspaces did not change the human perception while obscuring the machine perception of speaker attributes, indicating the inconsistency between the two perceptions. Moreover, the speaker-modified speech attained comparable WERs with the regenerated speech ($K_{\rm s}=0$), indicating that such alterations to the speaker embedding did not compromise the linguistic content.

In the residual subspace experiments, the WER significantly increased in the FAcodec model. Besides, a notable rise is found in the Diff-HierVC model, increasing from 4.17\% to 10.96\%. These observations suggest the influence of the residual subspace on linguistic content, demonstrating that the removal of this subspace is incapable of voice anonymization, which requires the preservation of linguistic content.

Seeing from the evaluations with the ECAPA-TDNN speaker extractor, comparing with the regenerated speech ($K_{\rm s}=0$), the removal of the secondary subspace \({\boldsymbol{S}}_{\rm s}=\{200, 45, -\}\) in FACodec achieved an increase in EER from 4.79\% to 8.76\%. Similarly, the subspace \({\boldsymbol{S}}_{\rm s}=\{218, 170, -\}\) in Diff-HierVC yielded an EER increase from 4.22\% to 6.65\%. Both modifications preserved human perception at the 100\% rate and did not cause degradation to the linguistic content. These results demonstrate that an asynchronous voice anonymization method can be developed by removing the contribution from the subspace in the speaker embedding. Particularly, in the FACodec model, it outperformed the adversarial approach \cite{wang2024async} at \(K_{\rm s}=45\), achieving a higher EER (8.76\% vs. 6.50\%) and a higher human perception preservation rate (100\% vs. 60.71\%).

\section{Conclusions and discussions}
\label{sec: conclusions}
This paper investigated the inconsistency between the machine and human perceptions on speaker attributes in the speech generation framework. It was conducted within the speaker variability subspaces of the speech generation models FACodec and Diff-HierVC. Experimental findings reveal that in both models, a subspace within the speaker embedding variability space exists, whereby the removal of its contribution from the speaker embedding alters machine-detectable speaker attributes while preserving human perception. Based on the investigation, an asynchronous voice anonymization method is developed through the removal of the subspace from the speaker embedding.

In addition to enhancing voice privacy protection, future research will focus on comprehensive evaluations of techniques for preserving speaker-independent attributes of the original speech, including speech quality and prosody, etc.

\printbibliography

\end{document}